\documentclass[11pt,dvips]{article}
\usepackage{epsfig,times} 
\usepackage{picinpar}
\usepackage{wrapfig}
\usepackage{floatflt}
\makeatletter
\renewcommand{\section}{\@startsection{section}{1}{0in}
        {0.4\baselineskip}{0.1\baselineskip}{\Large\bf}}
\renewcommand{\subsection}{\@startsection{subsection}{2}{0in}
        {0.25\baselineskip}{-\baselineskip}{\large\bf}}
\renewcommand{\subsubsection}{\@startsection{subsubsection}{3}{0in}
        {0.1\baselineskip}{-\baselineskip}{\normalsize\bf}}
\makeatother
\begin{document}

%
\thispagestyle{myheadings}
%
\markright{OG 3.3.01}
\begin{center}
%
{\LARGE \bf Are Galactic Gamma-Ray Bursters
the Main Source of Hadronic Non-Solar Cosmic Rays at all Energies ?}
\end{center}

\begin{center}
%
%
{\bf R. Plaga$^{1}$, O.C. de Jager$^{2}$, and A.Dar$^{3}$}\\
{\it $^{1}$ Max-Planck-Institut f\"ur Physik, 80805 M\"unchen, Germany\\
$^{2}$ Potchefstroomse Universiteit vir CHO, Potchefstroom 2520, South Africa\\
$^{3}$ Technion, Israel Institute of Technology, Haifa 32000, Israel}
\end{center}

\begin{center}
{\large \bf Abstract\\}
\end{center}
\vspace{-0.5ex}
%
%
\noindent
We propose a new hypothesis for the origin of non-solar
hadronic cosmic rays (CRs) at all energies:  Highly relativistic,
narrowly collimated jets from the birth or collapse of neutron
stars (NSs) in our Galaxy
accelerate ambient disk and halo matter to CR energies
and disperse it in ``hot spots'' which they form when they stop in the
Galactic halo. Such events - ``Galactic Gamma-Ray Bursters'' (GGRBs) - 
are proposed to cause cosmological gamma-ray bursts
(GRBs) in other galaxies when their beamed radiation happens to point
in our direction.
Our hypothesis naturally explains 
some observations which are difficult to understand with
the currently popular ideas about CR origin  -
e.g. the small Galacto-centric gradient of the cosmic-ray density
and the absence of the Greisen-Zatsepin-Kuzmin cutoff.
{\it Our idea stands or falls with the existence of the ``hot spots''
(``GGRB remnants'') in the Galactic halo.} We discuss their expected
observational signatures and find that they could appear as
EGRET unidentified high-latitude sources.              
%
\vspace{1ex}

%
%
\section{A fresh sheet of paper for the problem of CR origin}
\label{intro.sec}
\noindent
Neglecting solar modulation effects,
the flux of the dominating 
hadronic component of the local non-solar cosmic rays (CR),
as a function of energy, which has been measured between about a GeV and
3 $\cdot$ 10$^{20}$ eV, can be well described 
by one power-law which changes its slope slightly at only
two energies - at about 4 $\cdot$ 10$^{15}$ eV
(the ``knee'') and  3 $\cdot$ 10$^{18}$ eV (the ``ankle'').
This striking simplicity and unity of the data originally led
most authors to ascribe 
the origin of hadronic cosmic rays to a single source class,
either a Galactic one - such as supernova remnants (SNRs) - 
or extragalactic one, e.g. active galaxies (AGs).
Later an ``eclectic'' scenario for CR origin
became generally accepted: SNRs accelerate the CRs below the knee
and AGs are the source of CRs above the ankle.
This happened mainly because with a better
understanding of SNRs and AGs it became clear that it
is unlikely that either one can be the searched for
single source class.
Two consequences from this scenario were predicted
long ago:
\\
A. the analysis of $\gamma$-rays from the interaction of low energy
CRs with the interstellar medium in the Galaxy should show that
the CR density rapidly decreases with rising distance from the
Galactic centre, towards which SNRs are strongly concentrated
(Stecker \& Jones 1977).
\\
B. there should be an ``end of the the CR spectrum'' (Greisen 1966)
around 10$^{20}$ eV
- the Greisen-Zatsepin-Kuzmin (GZK) cutoff - because
of universal, i.e. not locally produced, CRs with higher energies
are severely depleted by interactions with the 3 K background
radiation.
\\
Both of these predictions have not been borne out by observations.
The data from three $\gamma$-ray satellites have consistently favoured
a much shallower Galacto-centric gradient 
of the hadronic cosmic-ray density than expected from SNRs.
In spite of great efforts
there is not a single {\it quantitative} description 
of the observed behaviour assuming the observed
Galacto-centric distribution of SNRs in the literature 
(Moskalenko \& Strong, 1998).
The CR spectrum definitely extends beyond the GZK cutoff 
(Takeda et al., 1998).
Moreover, up to now, there is no experimental indication of
{\it hadron} acceleration in SNRs. 
\\
This is motivation to search for
alternatives to the ``eclectic'' view.
At face value A. suggests that the hadronic CR sources
extend to the outskirts of our Galaxy, whereas B. means that UHE CRs
are produced near our Galaxy. This points towards
the Galactic halo as the {\it single} site of CR origin.
We propose that gamma-ray bursts that occur in our Galaxy
- i.e. a single source class - 
accelerate the observed hadronic CRs at this site.
The next section summarises our hypothesis
which has been described in detail
by Plaga \& Dar (1999), the other sections are new material.
\section{Plasmoid ejection in neutron star birth or collapse,
GRBs and CR origin}
\label{acc.sec}
\noindent 
Our two basic assumptions are:
\\
1. during {\it common} events in the
life of neutron stars (NSs) - such as their birth in a supernova
or their collapse to configuration with a different form
of nuclear matter (Dar 1999) - 
ultrarelativistic ``plasmoids'' (Lorentz factor $\Gamma$ $\approx$ 1000)
are emitted carrying
a significant fraction 
($\approx$ 10$^{52}$ ergs) of the NS's binding energy. 
After an initial expansion after the ejection they remain confined - 
by some as yet ill understood magnetic or inertial
(ram pressure) mechanism to a radius 
of $\approx$ 10$^{-3}$ pc. 
\\
2. after ejection the plasmoid decelerates - mainly via
accretion of ambient matter. Only after it slowed
down considerably, it begins a final expansion 
to about R$_f$ $\approx$ 0.1 pc and then dissipates
its kinetic energy, partly via particle acceleration.
\\
These assumptions seem natural because
ejection of plasmoids seems to be common
when mass is
accreted at a high rate onto a central compact object, a situation
expected in most scenarios
for the formation of  compact objects.
The Galactic superluminal source GRS 1915+105 
has properties consistent with our assumptions.
The ejection of plasmoids with a strong confinement 
mechanism obviously at work are commonly observed in 
extragalactic superluminal sources (there called ``plasmons'')
%
\begin{figwindow}[1,r,%
{\mbox{\epsfig{file=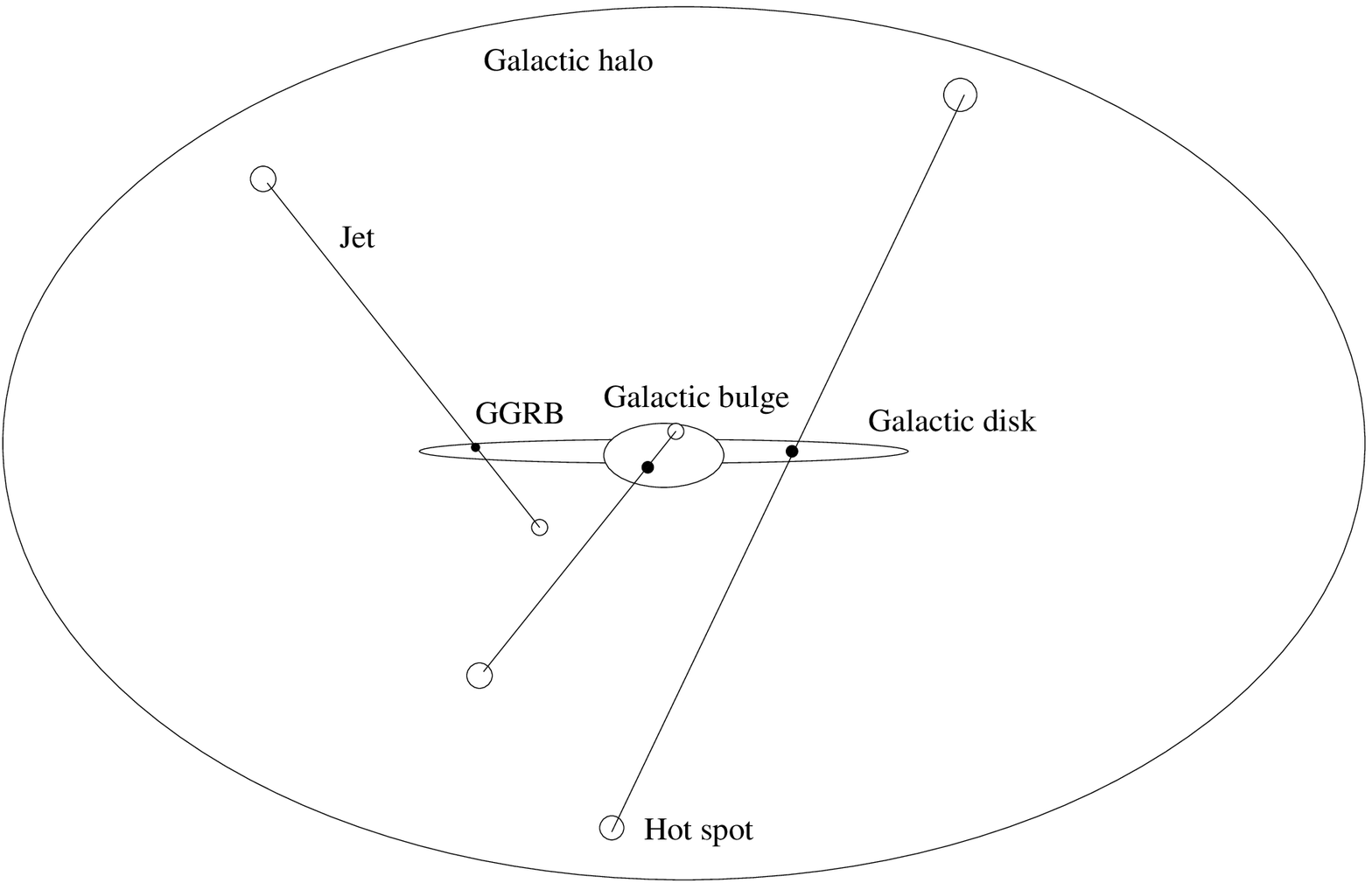,width=4.2in}}},%
{A schematic sketch of our scenario. The
birth or collapse of NSs in the disk of our Galaxy leads to an
ejection of two opposite jets (plasmoids) 
that produce ``hot spots'' when they stop in
an extended Galactic halo.}]
Synchrotron radiation from
decelerating plasmoids with properties
similar to the ones assumed here were shown to account for
GRB afterglows by Chiang and Dermer (1997).
In our scenario GRBs are expected to be
relativistically beamed by a factor 1/4$\Gamma^2$ $\approx$
2 $\times$ 10$^{-7}$ during the burst and initial afterglow phase: 
The observed GRB rate then  leads to a
GGRB rate similar to the NS birth rate of ca. one in 20 years.
\\
It can be shown that the plasmoid 
escapes into the Galactic halo if its final expansion does not
begin too early.
Because the final dissipational phase takes
place over a distance small compared to the total
travelled distance, the CR acceleration happens
in a small region which we call ``hot spots'' 
- in analogy to morphologically
similar regions in radio galaxies (fig.1).
Its equipartition magnetic field
is about 1 G.
{\it These hot spots are expected to accelerate particles
via shock-wave acceleration and magnetic reflection and
are identified as the single
source of hadronic CRs at all energies.}
Only about 1$\%$ of the plasmoid's kinetic energy
needs to be channelled into CR energy to explain 
its energy input into the Galaxy inferred from
spallation yields (Drury et al.,1989).
The plasmoids inject turbulent energy into the halo,
and in equipartition a halo magnetic field similar
to the one in disk is expected.
This leads to CR trapping in the halo.
If the plasmoids produce a hadronic CR power-law spectrum according to
E$^{-{\alpha}}$ with $\alpha$ = 2.2 below
and $\alpha$ = 2.5 above the knee, then - together
with an energy dependence of the interhalo diffusion coefficient
according to E$^{-0.5}$ - 
the complete observed CR spectrum from a GeV up to 10$^{20}$ eV
is naturally obtained.
\end{figwindow}
\section{The observational properties of the halo ``hot spots''}
\label{prop.sec}
\noindent
Our ignorance about the confinement mechanism of plasmoids
and relativistic jets in general introduces two 
basic unknowns which we shall leave as parameters in all expressions:
\\
{\it 1. the time scale $\tau_{\rm rad}$ 
of plasmoids' final expansion}, during which they dissipate most
of their kinetic energy into the form of CRs, radiation and
turbulent motion of the interhalo matter.
(Because of the strong beaming in our scenario
only a very small
fraction (less than 0.1 $\%$) of the initial kinetic
energy is radiated away in the prompt burst and early
afterglow.) 
With a particle density of 10$^{-3}$/cm$^3$ in the halo
the Lorentz factor of a plasmoid 
with a radius of 0.1 pc is decelerated to $\Gamma$=
2.5 in 10$^4$ years. Further
expansion will take place in the final phase of slow-down. 
Therefore we will specify $\tau_{\rm rad}$ in units of 1000 years 
below.
{\it 2. the distance d of a typical hot spot.}
The data on the Galacto-centric distribution of 
CRs suggest a scale of tens of kpc. 
\\
The hot spots bear some similarity to their
brethren in Fanaroff-Riley II galaxies, which
have been experimentally identified as prolific CR accelerators
(Rachen \& Biermann, 1993).  
Following them, 
the time scale for acceleration $\tau_+$ up to an energy E$_a$
in a jet with a speed/c = $\beta$
and a Kolmogorov turbulence spectrum is given as:
\begin{equation}
\tau_+  = 5 ({R_f/0.1 pc})^{2/3} 
\beta^{-2} ({({E_a/10^{20} eV})
/({B/1 G})})^{1/3} {\rm years}
\end{equation}
If this timescale (comfortably smaller than 
$\tau_{\rm rad}$)
is larger than the synchrotron cooling time
- this is certainly valid for electrons -
acceleration is limited by synchrotron
losses.
For protons diffusive loss could also be important.
de Jager et al. (1996) showed that the
maximal characteristic synchrotron energy 
for a particle with mass m is
given -independent of the magnetic field strength - as:
\begin{equation}
E_{\rm max} = 46.3 (m/m_{\rm proton}) {\rm GeV} 
\end{equation}
for maximal acceleration at a pitch angle of 90 $^\circ$.
The hot spot has to accelerate E$_{tot}$= 10$^{50}$ ergs in protons
to play its role as a universal CR source.
$\epsilon$ is the fraction of this energy emitted
as proton synchrotron radiation.
$\epsilon$ is smaller than one, its exact
value depends on the spectral index at the highest
energies and the importance of diffusive losses and
could only be precisely determined in a more detailed model.
We then obtain the luminosity of the hot spot
due to synchrotron radiation by requiring
$\int^{E_m}_0$ I$_{\rm syn}$ E dE = ($\epsilon {\rm E_{tot}})/ 
(\tau_{\rm rad} 4 \pi d^2$);
this leads to the following expression for the
hot-spot energy flux:
\begin{equation}
F_{HS}({\rm protons}) = 1.4 \times 10^{-8} \epsilon
({E/E_m})^{(-\alpha+3)/2} (1000 {\rm years}/\tau_{\rm rad})
(d/50 {\rm kpc})^{-2} 
{\rm erg~cm^{-2} sec^{-1}}
\end{equation}
The spectrum of the electron synchrotron has a spectral
index smaller by 0.5 above a synchrotron break energy
E$_{\rm break}$ $\approx$ 
8 $\left({(B/1G)\over(\tau_{\rm rad}/1000 y)}\right)$ keV,
which corresponds to a synchrotron break at 0.3 kHz.
This steeper spectrum leads to a dominance of proton
synchrotron radiation above about the IR region.
The potential importance of proton synchrotron radiation
in GRB afterglows (which are closely related to 
the hot spots in our scenario) 
was recognised by various
authors (e.g. Boettcher \& Dermer, 1998).
With our very simplifying assumptions a proton synchrotron
break could appear around 10$^{17}$ eV, but at these energies
diffusive losses from the hot spot - which counteract
against a break in the emitted charged particle spectrum - are important.
The exact synchrotron
spectrum in the $\gamma$-ray range then depends on the magnetic
fields around the hot spot. 
We assume that protons and electrons are accelerated 
to a given {\it Lorentz factor} with
equal efficiency by the relativistic 
shocks and bulk motions (Dar 1998).
Taking into account the synchrotron break
one then obtains for the radio intensity due to
electron synchrotron radiation:
\begin{equation}
I_{\rm HS}({\rm electrons}) = 6 
({\nu/5 {\rm Ghz}})^{-\alpha/2} (1000/\tau_{\rm rad})^{3/2}~{\rm mJy}
\label{esy}
\end{equation}
The proton synchrotron radiation contributes only about 0.1 mJy at
5 GHz.
\\
At energies above E$_{\rm max}$ (46 GeV)
the decay of $\pi_0$s
from proton interactions with ambient matter
swept up by the plasmoid
is expected to dominate the $\gamma$ radiation.
We estimate the integral flux 
above $\approx$ 10 GeV:
\begin{equation}
F_{\pi_0} = 2 \times 10^{-13} (\rho/50 {\rm cm}^{-3}) 
(d / 50 {\rm kpc})^{-2}
(E/{\rm TeV})^{-1.1} {\rm cm}^{-2} {\rm sec}^{-1}
\label{pga}
\end{equation}
This is on the order of 10 mCrab around one TeV, just
at the limit of a deep exposure with existing air-\v{C}erenkov
telescopes.
\section{Unidentified (UI) high-latitude 
$\gamma$ sources: ``hot spots'' in the Galactic halo?}
\label{uid.sec}
\noindent
The third EGRET catalogue lists 96 UI sources 
at high Galactic latitudes. 
Could some fraction of them be ``active'' hot spots?
As an example we discuss the brightest of them,
the UI source GRO J1835+5921 at a Galactic latitude
of 25 $^\circ$;
at energies above 1 GeV it is about half as 
bright as the Crab nebula (Nolan et al. 1996).
Its energy flux at 100 MeV is 1.1 $\times$ 10$^{-10}$ 
erg cm$^{-2}$ sec$^{-1}$
which is the one predicted above for a typical hot spot for $\epsilon$
$\approx$ 0.03. Its spectral 
index above 30 MeV is -1.69 $\pm$ 0.07 in
agreement with the index of -1.75 which follows for
$\alpha$=2.5 (the expected source index for UHE protons in
our scenario).
GRO J1835+5921 could be a (perhaps
exceptional) hot spot which proton synchrotron spectrum
extends up to E$_{\rm max}$.
In spite of the fact that its position is known with a precision
of 4.5 $^{\prime}$ no counterpart has been found
(the only object within its error box
is the IRAS source F18343+5913 which 
is misprinted in Nolan et al.(1996)
and identical to the main sequence K dwarf HD 172147;
hardly a $\gamma$-ray source).
The absence of a counterpart in the ROSAT 1 RXS RASS catalogue 
sets an upper limit of about 
4 $\times$ 10$^{-13}$ erg/cm$^2$/sec on emission
in the 0.1 - 2.4 keV band. This upper limit is about one order of magnitude
below the prediction with $\epsilon$=0.03
and $\alpha$=2.5, the predicted flux would be below
this limit for a CR source spectrum with $\alpha$=2. 
We therefore expect that a search for X-ray emission
near GRO J1835+5921 will detect a hard X-ray source near
the sensitivity limit of the RASS.
The absence of a radio source in the 87GB (TXS)
catalogue sets upper limits of 25 (400) mJy at 4.85 (0.365) GHz.
These values are above the expected one from eq. \ref{esy}.
A deeper search near GRO J1835+5921 
should reveal a soft nonthermal radio source 
(spectral index -1) with an angular 
extension {\it and proper motion per year}
of a few tenths of an arcsec. 
The clearest evidence in favour of an identification 
of GRO J1835+5921 as a  hot spot
- rather than as a very nearby Geminga like radio-quite
pulsar (Mukherjee et al., 1995) or a flaring star (\"Ozel
\& Thompson, 1996) -
would be the detection of VHE energy $\gamma$-radiation
at energies well above the range thought likely for pulsar
emission, say above 100 GeV (eq. \ref{pga}).
The hot spots could also have spectral
synchrotron breaks below E$_{\rm max}$ and could then be 
identified with softer high-latitude UI sources.
This work made use of the Nasa Extragalactic Database (NED), for
references of catalogues refer to this source.
%
%
%
%
%
%
\vspace{1ex}
\begin{center}
{\Large\bf References}
\end{center}
%
Boettcher, M. \& Dermer, C.D. 1998, ApJ 499, L133\\
Chiang, J. \& Dermer, C.D., 1997, astro-ph/9708035\\
de Jager, O. et al. 1996, ApJ 457, 253\\
Dar, A. 1998, astro-ph/9809163\\
Dar, A. 1999, astro-ph/9902017, Proc. Rome GRB Conf, A \& A in press\\
Dar, A. \& Plaga, R. 1999, astro-ph/9902138, A \& A in press\\
Drury, L. O'C., Markiewicz, W. J. \& V\"olk, H. J. 1989, A \& A 225, 179\\
Greisen, K. 1966, PRL 16, 748\\
Moskalenko, I.V. \& Strong, A.W. 1998, ApJ 509, 212\\
Mukherjee, R. et al. 1995, ApJ 441, L61\\
Nolan, P.L. et al. 1996, ApJ 459, 100\\
\"Ozel, M.E. \& Thompson, D.J. 1996, ApJ 463, 105 \\ 
Rachen, J.P. \& Biermann, P.L. 1993, A \& A 272, 161\\
Stecker, F.W. \& Jones, F.C. 1977, ApJ 217, 843\\ 
Takeda, M. et al. 1998, PRL 81, 1163
\end{document}